\documentclass[aps,prb,twocolumn,showpacs]{revtex4-1}

\usepackage{amsmath}
\usepackage{amssymb}
\usepackage{graphicx}
\usepackage{bm}

\begin{document}
\title{Majorana-assisted nonlocal electron transport through a floating
topological superconductor}
\author{Jascha Ulrich}
\author{Fabian Hassler}

\affiliation{JARA-Institute for Quantum Information, RWTH Aachen, D-52074,
  Germany}

\date{June 2015}

\begin{abstract}
The nonlocal nature of the fermionic mode spanned by a pair of Majorana bound
states in a one-dimensional topological superconductor has inspired many
proposals aiming at demonstrating this property in transport.  In particular,
transport through the mode from a lead attached to the left bound state to a
lead attached to the right will result in current cross-correlations.  For
ideal zero modes on a grounded superconductor, the cross-correlations are
however completely suppressed in favor of purely local Andreev reflection.  In
order to obtain a non-vanishing cross-correlation, previous studies have
required the presence of an additional global charging energy.  Adding
nonlocal terms in the form of a global charging energy to the Hamiltonian
when testing the intrinsic nonlocality of the Majorana modes seems to be
conceptually troublesome.  Here, we show that a floating superconductor allows
to observe nonlocal current correlations in the absence of charging energy.
We show that the non-interacting and the Coulomb-blockade regime have the same
peak conductance $e^2/h$ but different shot-noise power; while the shot noise
is sub-Poissonian in the Coulomb-blockade regime in the large bias limit,
Poissonian shot noise is generically obtained in the non-interacting case.

\end{abstract}

\maketitle 

\section{Introduction}
A pair of Majorana bound states at the ends of one-dimensional \emph{p}-wave
superconductors hosts a nonlocal fermionic mode at an energy close to the
middle of the energy gap.\cite{kitaev:01} The
nonlocality of the fermionic mode has inspired proposals to observe nonlocal
current correlations which arise when electrons are transported through the
mode from a lead contacting the left bound state to a lead contacting the
right.\cite{semenoff:07} Unfortunately, nonlocal transport is completely
suppressed in favor of local Andreev reflection for ideal Majorana bound
states at zero energy and the currents through the left and right lead are
uncorrelated with characteristic peak conductances $2e^2/h$. \cite{bolech:07}
Later, it has been pointed out that a coupling $t$ between the Majorana end
states that is larger then the lead-induced level broadening $\Gamma$ is
required for a recovery of nonlocal current correlations.\cite{nilsson:08,
tewari:08, liu:13, zocher:13} By increasing the length of the
superconductor $L$ the coupling constant $t$ decays exponentially such that for
sufficiently long topological superconductors the transport remains local
despite the nonlocal nature of the fermionic zero mode.

It was first shown in Ref.~\onlinecite{fu:10} that to observe nonlocal current
cross-correlations for ideal Majorana zero modes with $t=0$, it is important
that the superconductor carrying the Majorana zero modes is disconnected from
the ground plate, resulting in a finite charging energy.  In the Coulomb
blockade regime, local Andreev reflection is blocked energetically and
nonlocal transport with a system size independent level broadening $\Gamma$
and a peak conductance of $e^2/h$ at zero temperature is obtained.  Due to the
suppression of local Andreev reflection, the analysis can be carried out in
the usual mean-field setup of superconductivity without violating current
conservation.\cite{anantram:96} Subsequent studies have considered the
behavior of the peak conductance as the charging energy is lowered away from
the Coulomb blockade regime, finding a smooth increase from the peak
conductance $e^2/h$ in the Coulomb blockade regime to $2e^2/h$ in the absence
of charging energy.\cite{hutzen:12, zazunov:11}

\begin{figure}[tb]
\centering
\includegraphics{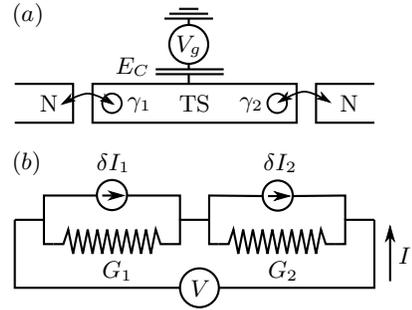}
\caption{%
($a$) Setup consisting of a one-dimensional topological superconductor (TS)
that is floating and hosts ideal zero-energy Majorana bound states $\gamma_1$,
$\gamma_2$ contacted by two normal-conducting, non-interacting leads (N).  The
superconductor is subject to a charging energy $E_C$ such that a preferred
number of charges is controllable through a gate voltage $V_g$. We study the
system in the case $E_C=0$ and compare it to known results for large $E_C$.
($b$) In absence of charging energy, $E_C = 0$, the conductance and shot noise
of the system can be understood in terms of an equivalent circuit consisting
of (noiseless) conductances $G_i$ which are associated with local Andreev
reflection off the bound state $\gamma_i$ and current sources generating noise
currents $\delta I_i$.
}
\label{fig:setup}
\end{figure}

The system size independence of $\Gamma$ in Ref.~\onlinecite{fu:10} must be
contrasted with the transmission of spinless electrons through a double
barrier, where the level broadening $\Gamma$ scales with system size as
$L^{-1}$ such that the conductance quantization at a value $e^2/h$ is lost for
any finite temperature in the infinite system limit.\cite{nazarov} In
contrast, the peak conductance $e^2/h$ survives at finite temperatures in the
Majorana case, justifying the terminology of calling it `\emph{nonlocal
transport}'.  However, the presence of the large global charging energy, which
is a strong, nonlocal perturbation at the level of the Hamiltonian, makes it
difficult to unambiguously attribute the nonlocal current correlations to the
presence of the Majorana bound states.\cite{sau:12} In this sense, a large
charging energy can be considered ill-suited for probing the Majorana-induced
nonlocality.

In this work, we consider a floating superconductor where the charging energy
$E_C$ vanishes.  We will contrast the results to the well-studied limit of
Coulomb-blockade $E_C \to \infty$ where the model maps onto a resonant-level
model.\cite{fu:10} We show that consideration of a floating topological
superconductor in a setup as depicted in Fig.~\ref{fig:setup}($a$) allows to
observe nonlocal current correlations for ideal Majorana bound states at zero
energy in the absence of charging effects.  Interestingly, we find that the
peak conductance is given by $e^2/h$ both in the non-interacting and in the
Coulomb blockade limit.  The contradiction to the results of
Refs.~\onlinecite{hutzen:12, zazunov:11} can be resolved by noting that away
from the Coulomb blockade regime, transport into the Cooper pair condensate is
no longer blocked energetically and current conservation has to be enforced
self-consistently.\cite{anantram:96} This appears to be technically demanding
in the interacting setup of finite charging energy.  Even though the peak
conductances does not distinguish the two regimes, the transport mechanisms
differ strongly.  In the Coulomb-blockade regime, the Cooper-pair condensate
is irrelevant and transport proceeds by sequential tunneling which can be
modeled as coherent transport through a double barrier.\cite{blanter:00,
chen:92, davies:92} In contrast, in the non-interacting case, transport is
incoherent due to the presence of the Cooper-pair condensate which acts as a
phase-breaking scatterer.\cite{datta} We will show that while the difference
between the two transport regimes is not captured by the peak conductance, it
is reflected in the shot noise properties.

The effective phase-breaking in electron transport due to the Cooper-pair
condensate allows to understand the transport in terms of an equivalent,
classical circuit depicted in Fig.~\ref{fig:setup}($b$). In this circuit,
Andreev reflection events at the left and the right terminal are modeled by
noiseless conductances $G_1$, $G_2$ and their corresponding noise currents
$\delta I_i$ are generated by current sources in parallel to the conductances.
Straightforward application of Kirchhoff rules yields the total  conductance
$G= \langle I\rangle/V = G_1 G_2/(G_1 + G_2)$ since the resistances $G_i^{-1}$
simply add up.  In presence of ideal Majorana bound states, the each of the
two conductances  $G_1$ and $G_2$ in series peak at a value equal to the
conductance quantum $2e^2/h$. Accordingly, the total conductance $G$ has a
peak vale of half a conductance quantum $e^2/h$.  Similarly, assuming
uncorrelated noise currents $\langle \delta I_1(t) \delta I_2(t') \rangle =
0$, one obtains the total current noise-power  $S= \int\!dt\, \langle \delta
I(0) \delta I(t) \rangle =  (G_2^2 S_{1}^{(0)} + G_1^2 S_{2}^{(0)}) /(G_1 +
G_2)^2 $ with $S_{j}^{(0)} = \int \!dt\,\langle \delta I_j(0) \delta I_j(t)
\rangle$;  here, the prefactors in front of the factors $S_i^{(0)}$ originate
from current division and give the (squared) fraction of the noise currents
$\delta I_i$ contributing to the total current $I$.  The above results
correspond directly to our results Eqs.~\eqref{eq:conductance}
and~\eqref{eq:noise}, yielding an appealingly simple picture of transport
through the system.  By analyzing the interplay between the bias dependencies
of the conductances and the noise currents, we will show that the shot noise
in the non-interacting limit $E_C = 0$ generically becomes  Poissonian in the
large-bias limit.  In contrast, the shot noise is sub-Poissonian in the
Coulomb-blockade regime.

The outline of the paper is as follows.  We use Sec.~\ref{sec:grounded} to fix
our notation and give a brief review of results obtained in previous studies
for the grounded case.  We discuss the conductance and shot noise obtained in
the floating case for $E_C =0$ in Sec.~\ref{sec:floating}.  We will compare
the results to the well-studied case of a superconducting island in the
Coulomb-blockade regime in Sec.~\ref{sec:cmp_block} and finish with a
discussion of our results.

\section{Review of the grounded case}
\label{sec:grounded}
Our system of interest is a one-dimensional $p$-wave spinless superconductor
that hosts a pair of Majorana bound states $\gamma_i$, $i=1,2$,  each of them
in tunnel contact with a single normal-conducting lead $i$ as depicted in
Fig.~\ref{fig:setup}($a$).  Before discussing the case of a floating
superconductor that is disconnected from the ground, we start with a review of
the grounded case.  For the grounded case, we describe the system using a
mean-field Bogoliubov-de-Gennes (BdG) Hamiltonian and calculate the transport
properties using a scattering approach.  The entries $s_{ij}^{\alpha \beta}$
of the scattering matrix describe the transport of particle of type $\beta$
from contact $j$ to contact $i$ as a particle of type $\alpha$, where $i,j \in
\{1, 2, S\}$ can be either of the normal-conducting terminals or the
superconductor ($S$) and $\alpha$, $\beta \in \{e, h\}$ denote electrons and
holes.  With the transfer probabilities $T_{ij}^{\alpha \beta} = |
s_{ij}^{\alpha \beta}|^2$, the steady-state current at the normal-conducting
terminal $i$ into the system can be written as \cite{lambert:93, anantram:96}
\begin{align}\label{eq:current_general} I_i = \frac{e}{h} \int d E \sum_{j
\neq S} \bigl( \delta_{ij} - T_{ij}^{ee} + T_{ij}^{he} \bigr) \bigl( f_{j,E} -
f_{S,E}\bigr) \end{align}
with $f_{j,E} = \Theta(\mu_j -E)$ the Fermi distribution of contact $j$ at an
electro-chemical potential $\mu_j$ at zero temperatures with $\Theta(x)$ the
unit step function.  In the following, we choose a gauge where the energy $E$
is measured with respect to the chemical potential $\mu_S$ of the
superconductor such that $\mu_j = eV_j$ with $V_j$ denoting the voltage
difference between the lead $j$ and the superconductor.  The transport
properties are described by the differential conductances $G_{ij} = \partial
I_i / \partial V_j$ which have the form
\begin{align}\label{eq:conductance_general}
  G_{ij} = \frac{\partial I_i}{\partial V_j} = \frac{e^2}h \bigl[
\delta_{ij} - T_{ij}^{ee}(eV_j) + T_{ij}^{he}(eV_j)\bigr]
\end{align}
at zero temperature; here, $T_{ij}^{\alpha \beta}(eV)$ denotes the
transmission probability at energy $eV$.  For ideal Majorana bound states and
low voltages with energy below the gap of the superconductor, contributions
from crossed Andreev reflection ($T_{12}^{he}$), normal transmission
($T_{12}^{ee}$), and quasiparticle transport into the supercondutor
($T_{jS}^{\alpha\beta}$) vanish.\cite{flensberg:10b} Consequently, the only
processes which are possible are normal-  ($T^{ee}_{jj}$) and
Andreev-reflection ($T^{he}_{jj}$) such that the quasiparticle conservation
leads to the sum rule $T_{jj}^{ee} + T_{jj}^{he} = 1$.  As a result, the
conductance matrix becomes local, $G_{ij} = \delta_{ij} G_j$, with
\begin{align}\label{eq:conductance_ideal_majorana}
  G_{j} =  \frac{2e^2}h T_{jj}^{he}(eV_j).
\end{align} 
As shown in App.~\ref{sec:smatrix_ns}, explicit calculations of the
scattering matrix for ideal Majorana bound states yield $T_{jj}^{he}(0) = 1$,
implying resonant Andreev reflection with a peak conductance of $2e^2/h$ at
zero bias $V_j = 0$.  \cite{flensberg:10b}

The nonlocal current correlations can be investigated through the
zero-frequency shot noise
\begin{align} \label{eq:noise_general} 
S_{ij} = \int d t \, \Bigl[\langle 
\hat I_{i}(0)  \hat I_{j}(t)  \rangle -\langle \hat I\rangle^2 \Bigr], 
\end{align}
defined in terms of the current operator $\hat I_i(t)$ at terminal $i$.  For
ideal Majorana bound states in a three-terminal NSN setup where the
superconductor is grounded, the shot noise at each terminal takes the same
form as in a conventional NS junction $S_{ij}^{(0)} = \delta_{ij} S_{j}^{(0)}$
with \cite{dejong:94}
\begin{align}\label{eq:noise_ns}
 S_{j}^{(0)}= \frac{4e^2}{h} \int d E \, T_{jj}^{he}(1-T_{jj}^{he}) 
( f_{j,E} - f_{S,E})^2,
\end{align}
and no current cross-correlation is present. \cite{bolech:07}

\section{Floating case without interactions}
\label{sec:floating}
In the case of the floating superconductor without interactions $E_C=0$, the
mean-field BdG formalism which neglects the Cooper-pair condensate dynamics
cannot be applied naively since electron pairs can be converted into Cooper
pairs, resulting in a net current $I_S = -(I_1 + I_2)$ out of the
superconducting reservoir.  For a floating superconductor in the stationary
regime, no net current $I_S$ can be drawn from the reservoir.\cite{lambert:93,
anantram:96} The condition $I_S = 0$ can always be fulfilled by varying the
sum of the voltage $V_1 + V_2$ (which corresponds to changing the chemical
potential of the superconductors with respect to the leads) while keeping the
voltage difference $V = V_1 - V_2$ between the two terminals fixed. The
superconducting reservoir then serves as a voltage probe breaking the phase
coherence of the quasiparticles without affecting the current balance.

For a general floating NSN setup, one obtains the
conductance\cite{lambert:93, anantram:96}
\begin{align} 
G = \frac{\partial I_1}{\partial V}\Big\vert_{I_S=0} = \frac{
G_{11} G_{22} - G_{12} G_{21} }{
G_{11} +
G_{12} + G_{21} + G_{22}},
\end{align}
where all the conductances $G_{ij}$ have to be evaluated at the voltages $V_i$
corresponding to $I_S =0$. For ideal Majorana bound
states, we have $G_{12} = G_{21} = 0$ and the formula simplifies to
\begin{align}\label{eq:conductance}
G = \frac{G_{1}G_{2}}{G_1 + G_2},
\end{align}
which corresponds directly to the conductance that we have derived in the
introduction for the equivalent circuit Fig.~\ref{fig:setup}($b$).  As
previously explained, the total resistance $G^{-1}$ is simply determined by
the sum of the resistances $G_{1}^{-1}$, $G_{2}^{-1}$ associated with Andreev
reflection processes at the left and right terminal and the total conductance
$G$ peaks when the conductances $G_1$, $G_2$ attain their maximum values at
$V_1 = V_2 = 0$.  At this point, current conservation follows trivially from
$I_1 = I_2 = 0$ and one obtains a peak conductance of $e^2/h$ at zero bias
$V=0$.  The peak conductance in the floating case is thus precisely one half
of the peak conductance expected for a grounded superconductor.

For the calculation of the noise in the floating case, we follow the approach
of Ref.~\onlinecite{anantram:96} which assumes that the current fluctuations
can be modeled as Langevin forces $\delta \hat I_j$ that drive the
fluctuations $-e \delta V$ in the chemical potential of the superconductor
which in our gauge corresponds to modifying the voltages according to $\delta V_{1/2} = \delta V$.  Although Ref.~\onlinecite{anantram:96} presents the results
in linear response in the voltages $V_j$ where they can be rigorously
justified through fluctuation-dissipation relations,\cite{landau:9} the
treatment formally only requires linear response in the fluctuations $\delta
V$ of the voltages $V_j$.  This can be seen by expanding the zero-frequency
current operator $\hat I_i$ of lead $i$ in the form
\begin{align}\label{eq:current_langevin}
  \hat I_i = I_i + \sum_{j\neq S}G_{ij}\, \delta V + \delta \hat I_i, 
\end{align}
where the first term is just the average current
Eq.~\eqref{eq:current_general} and $\delta \hat I_i$ are the Langevin forces
defined by $\langle \delta \hat I_i(t) \rangle = 0$, $\int\!dt\,\langle
\delta\hat I_i(0) \delta \hat I_j(t) \rangle = S_{ij}^{(0)}$ with the shot
noise $S_{ij}^{(0)}$ of the grounded case.  In
equation~\eqref{eq:current_langevin}, the expressions $I_i$, $G_{ij}$ and
$S_{ij}^{(0)}$ all have to be evaluated at the voltages corresponding to
$I_S=0$.  For long measurement times which corresponds to low
frequencies, current conservation has to hold not only on average but as an
operator identity $\sum_i \hat I_i = 0$. \cite{anantram:96}

Using the current conservation, the result of Ref.~\onlinecite{anantram:96}
for the shot noise in a floating two-terminal setup is recovered. The shot
noise is maximally cross-correlated with $S_{11} = S_{22} = - S_{21} = -S_{12}
=S$ where
\begin{align}\label{eq:noise}
S 
= \frac{G_{2}^2 S_{1}^{(0)} +  G_{1}^2 S_{2}^{(0)}}{(G_{1}+G_{2})^2}.
\end{align}
Note that the expression of the shot noise \eqref{eq:noise} can directly be
understood in terms of the equivalent circuit Fig.~\ref{fig:setup}($b$) that
was discussed in the introduction. However, in order to determine the concrete
bias dependence of the noise \eqref{eq:noise}, we need to solve the
self-consistency equation $I_S=0$ determining the voltages $V_i$ given the
bias $V$. We determine this relation using a microscopic model of two ideal
Majorana bound states $\gamma_1$, $\gamma_2$ that are each connected to a
lead. Because there is no coupling between the ideal Majorana modes, the
transport is  local and we can use the results of App.~\ref{sec:smatrix_ns} in
the expressions Eqs.~\eqref{eq:current_general},
\eqref{eq:conductance_ideal_majorana} and \eqref{eq:noise_ns} for $I_{j}$,
$G_{j}$, and $S^{(0)}_{j}$. We obtain
\begin{align}
  G_j &= \frac{2 e^2}h (1 + v_j^2), \qquad
I_j = \frac{4 e}h \Gamma_j \arctan(v_j), \label{eq:current_microscopic}\\
S_{j}^{(0)} &= \frac{4e^2}h \Gamma_j \bigl( \arctan |v_j| - |v_j|/(1+v_j^2)\bigr),
\label{eq:shot_noise_microscopic}
\end{align}
with $v_j = e V_j/2 \Gamma_j$ where we have introduced the coupling strength
$\Gamma_j$ of the Majorana fermion $\gamma_j$ to the nearby lead.

For the following, a possible asymmetry between the lead couplings $\Gamma_j$
will  be important. Thus, we will label the leads by the indices $j \in
\{\text{min}, \text{max}\}$ such that $\Gamma_\text{min} \leq
\Gamma_\text{max}$. Note, however, that for the voltage drop $V_\text{min}$,
the label `min' merely means that the voltage drops at the terminal
with the smaller lead coupling $\Gamma_\text{min}$ which implies that
$V_\text{min} \geq V_\text{max}$ as we will show below.

Solving the constraint $I_S = 0$ valid for a floating superconducting in terms
of $v_\text{max}$ yields
\begin{align}\label{eq:voltage_constraint}
v_\text{max} = - \tan\bigl[\Gamma_\text{min}
\arctan(v_\text{min})/\Gamma_\text{max}\bigr].
\end{align}
\begin{figure}[tb] 
\centering \includegraphics{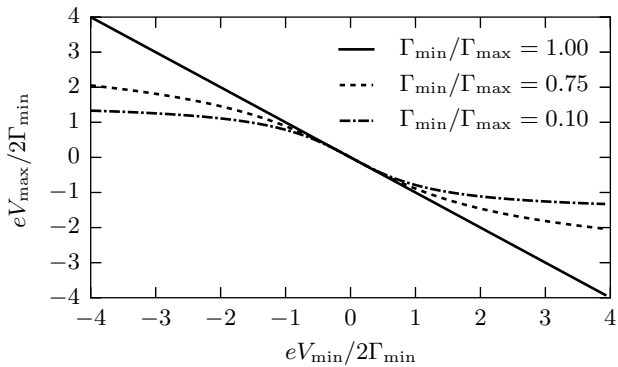}
\caption{Relation between the voltage drop $V_\text{min}$ at the terminal with
the smaller lead coupling $\Gamma_\text{min}$ and the voltage drop
$V_\text{max}$ at the terminal with the larger lead coupling
$\Gamma_\text{max}$ for a floating superconductor.}
\label{fig:voltages} 
\end{figure}%
The physics behind the transport can be entirely understood through the
voltage relation \eqref{eq:voltage_constraint}, which is plotted in
Fig.~\ref{fig:voltages} for different ratios
$\Gamma_\text{min}/\Gamma_\text{max}$.  We find that for small applied bias
voltages $V \ll V_- = 2 \Gamma_\text{min}/e$ the voltage drop is symmetric
with $|V_\text{max}| =  |V_\text{min}| =\tfrac12 V$. This results corresponds
to the linearized regime in Fig.~\ref{fig:voltages}. Note that for the special
case of symmetric couplings this relation holds exactly for arbitrary
voltages.  At larger voltage bias, the voltage $V_\text{max}$ saturates at a
value
\[
  V_\text{sat} = 2 \Gamma_1 \tan(\pi \Gamma_\text{min}/2
\Gamma_\text{max}).
\]
The origin of the saturation is the effect that a single Andreev level can
carry at most a current $2\pi e \Gamma/h$.  The total current through the
device is thus limited by $2\pi e \Gamma_\text{min}/h$.  In order for $I_S=0$
to hold, the current through the junction with the coupling
$\Gamma_\text{max}$ must not exceed this value.  Using
\eqref{eq:current_microscopic}, with $I_\text{max} = 2\pi e
\Gamma_\text{min}/h$, we obtain that the voltage $V_\text{max}$ has to
saturate at the value $V_\text{sat}$.  If the bias voltage exceeds this
saturation value almost all the voltage drop will be across the junction with
the smaller coupling, i.e., we have $V= V_\text{min}$ for $V \gg
V_\text{sat}$.  We highlight that the existence of this regime is a
consequence of the resonant structure of Andreev reflection at the NS
interfaces.

\begin{figure}[tb]
\centering
\includegraphics{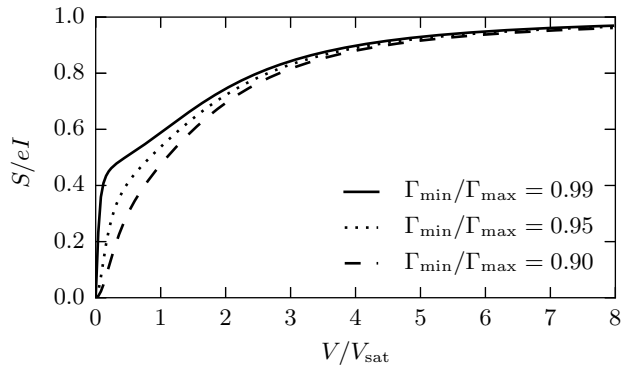}
\caption{Fano factor $F = S/eI$ in the floating system as a function of the
  voltage $V/V_\text{sat}$, where $e V_\text{sat} = 2 \Gamma_\text{max}
\tan(\Gamma_\text{min} \pi/2 \Gamma_\text{max})$ is the largest voltage
$V_\text{max}$ that can drop at the terminal with the larger lead coupling
$\Gamma_\text{max}$.  For voltages $V \gg V_\text{sat}$, nearly all the
voltage drops at the terminal with the smaller lead coupling
$\Gamma_\text{min}$ and both the shot noise $S/e$ and the current $I$ tend to
the maximal current $2 \pi e \Gamma_\text{min}/h$ that can be driven through
the system.  Consequently, for voltages $V \gg V_\text{sat}$, the Fano factor
reaches one.}
\label{fig:fano_factor_vplus}
\end{figure}

From the behavior of the voltages, all the other transport characteristics
follow.  In the regime of low bias $V \ll V_-$, the relation have
$V_\text{min} = - V_\text{max}$ leads to a non-trivial relation between the
conductances $G_\text{min}/G_\text{max} = 1 - (\Gamma_\text{min}^2 +
\Gamma_\text{max}^2) e^2 V^2/ 16 \Gamma_\text{min}^2 \Gamma_\text{max}^2$ and
the behavior of the shot noise in the floating case is determined both by the
bias dependence of the conductances and the shot noises $S_\text{min}^{(0)}$,
$S_\text{max}^{(0)}$.  For $V \ll V_-$, we obtain a Fano factor $F = S/eI= (1
+ \Gamma_\text{min}^2/\Gamma_\text{max}^2) V^2/24 V_-^2 \ll 1$ which indicates
sub-Poissonian noise. For $V\to 0$, we obtain the result $F=0$ consistent with
the fact that the local transport is carried by a  resonant level with unit
transmission probability implying a vanishing Fano factor due to the Pauli
exclusion principle.\cite{blanter:00}

Since for $V \gg V_\text{sat}$, the voltage drop $V_\text{min}$ at the
terminal with the smaller coupling $\Gamma_\text{min}$ is much larger than the
voltage drop $V_\text{max}$ at the other terminal, the conductances in that
regime obey $G_\text{min} \ll G_\text{max}$.  Consequently, we find $S =
S_\text{min}^{(0)}$ which reflects the fact that according to the circuit
Fig.~\ref{fig:setup}($b$), it is the noise current generated at the terminal
associated with $\Gamma_\text{min}$ which flows through the larger conductance
and thus dominates the shot noise for $V \gg V_\text{sat}$.  Using the
explicit form of the noise \eqref{eq:shot_noise_microscopic} and the current
\eqref{eq:current_microscopic}, one finds that the Fano factor $F = S/eI$
tends to one for $V \gg V_\text{sat}$.  This behavior is illustrated in
Fig.~\ref{fig:fano_factor_vplus} for different ratios
$\Gamma_\text{min}/\Gamma_\text{max}$ close to one.

\section{Comparison to the Coulomb-blockade case}\label{sec:cmp_block}

The behavior of the shot noise power \eqref{eq:noise} in the non-interacting
case $E_C = 0$ must be contrasted with the shot noise power obtained for the
Coulomb-blockade regime $E_C \rightarrow \infty$ of Ref.~\onlinecite{fu:10}.
In the case where only two charge states are important, transport proceeds by
sequential tunneling and can be mapped onto transport though a double barrier
where tunneling through the barriers occurs at a rate
$\Gamma_\text{min}/\hbar$, $\Gamma_\text{max}/\hbar$.  \cite{davies:92}
Current and shot noise in that case can be obtained using the classical
circuit Fig.~\ref{fig:setup}($b$) and its corresponding shot noise expression
\eqref{eq:noise} if the conductances $G_\text{min}$, $G_\text{max}$ model the
transmission through a single barrier.\cite{blanter:00} In this case, the
conductances $G_\text{min}$, $G_\text{max}$ are simply proportional to the
couplings $\Gamma_\text{min}$, $\Gamma_\text{max}$ which are
energy-independent for wide leads, showing that the voltage drops
$V_\text{min}$, $V_\text{max}$ at the left and the right barrier must be
distributed according to $V_\text{min}/V_\text{max} =
\Gamma_\text{max}/\Gamma_\text{min}$.  Since neither of the voltage drops
saturates, both the Fano factors associated with transmission through the left
and right barrier $S_1^{(0)}/e I$, $S_2^{(0)}/e I$ will tend to one in the
large bias regime $eV \gg \Gamma_\text{min}, \Gamma_\text{max}$ and one
obtains the limiting expression $F_\text{db} = (\Gamma_\text{min}^2 +
\Gamma_\text{max}^2)/(\Gamma_\text{min} + \Gamma_\text{max})^2 < 1$ for the
Fano factor of the double barrier in the large bias limit.\cite{chen:92,
davies:92, blanter:00} This shows that the Fano factor is suppressed as
compared to the single barrier case with a maximal suppression of $1/2$
obtained for symmetric coupling $\Gamma_\text{min} = \Gamma_\text{max}$.

\begin{figure}[tb] 
\centering \includegraphics{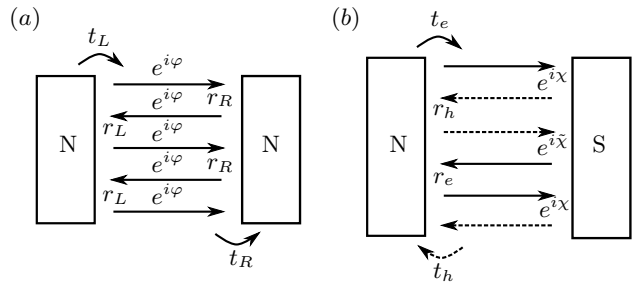}
\caption{This plot shows the equivalence of the tunneling through a symmetric
  double barrier to the Andreev reflection off a NS interface.  ($a$) The
  tunneling amplitudes for a double barrier
$t=\sum_{m=0}^\infty A_m$ is a sum of the amplitudes $A_m = t_L e^{i \varphi}
(r_R e^{2i \varphi} r_L)^m t_R$ for $m$ round trips between the barriers with
the amplitudes $r_{L/R}$ and $t_{L/R}$ of reflection and transmission at the
left/right barrier and the phase $e^{i \varphi}$ accumulated during one
traversal of the barrier region.  ($b$) The tunneling amplitude for Andreev
reflection $t^{eh}$ at a NS interface is similarly given by the sum of the
amplitude $A_m = t_e e^{i\chi} (r_h e^{i\tilde \chi} r_h e^{i \chi})^m t_h$
for $m$ round trips with the amplitudes $r_{e/h}$ and $t_{e/h}$ for reflection
and transmission of an electron/hole at the tunneling barrier and the
amplitudes $e^{i \chi}$, $e^{i \tilde \chi}$ for Andreev reflection as a hole
or electron at the superconductor.  For energies much smaller than the gap,
the particle-hole symmetry dictates that $\chi + \tilde \chi = -\pi$.  Thus,
we obtain the result\cite{nazarov} that Andreev reflection can be understood
as transmission through a symmetric double barrier with $T_L = |t_L|^2 = T_R =
|t_R|^2$ if we identify $\varphi = -\pi/2$ and $t_{e/h} = t_{L/R}$, $r_{e/h} =
r_{L/R}$.}
\label{fig:double_barrier} 
\end{figure}

In the following, we want to give an intuitive picture for the reduction of
the Fano factor in the Coulomb-blockade regime with respect to the case of a
noninteracting floating superconductor with $E_C =0$. In the former case, the
Fano factor is given by the generic result $F_\text{db}$ for an asymmetric
double barrier which is smaller than one. In the latter case, for large bias,
the voltage generically drops solely over the junction with coupling
$\Gamma_\text{min}$ and the shot noise as well as the current can be evaluated
in an NS-setup involving only this junction. The transport through the NS
junction involves Andreev reflections. It is a well-known fact that the
Andreev reflection processes in an NS-junction can also be mapped onto the
problem of transmission of electrons through a symmetric double barrier with
the concreted identification explained in Fig.~\ref{fig:double_barrier}. It is
however important to realize that the elementary processes in the transport
involve Cooper pairs with charge $2e$ such that the Fano factor is a twice
large than the one for the symmetric double barrier. Given the expression
$F_\text{db}$ above, we thus obtain the result that the Fano factor of the
noninteracting setup tends to one for large bias.

\section{Conclusions}
In this work, we have studied two-terminal transport between two
normal-conducting metallic leads contacting the Majorana bound states of a
floating topological superconductor in absence of charging energy.  The
nonlocal current correlations reflecting Majorana-assisted nonlocal electron
transport that were previously shown for the Coulomb-blockade
regime\cite{fu:10} are also obtained in the non-interacting case.  The
presence of the phase-breaking Cooper-pair condensate allows modeling the
transport processes as an incoherent combination of Andreev reflection events
at the left and the right NS junction.  Using this model, we have shown that
the peak conductance $e^2/h$ previously obtained for the Coulomb-blockade
regime also follows in the non-interacting case as a simple consequence of
current division.  We have shown that the difference between the
non-interacting case and the Coulomb-blockade regime is reflected in the shot
noise power, which generically becomes Poissonian in the high bias limit of
the non-interacting case, whereas it stays sub-Poissonian in the
Coulomb-blockade regime.  We thus find that the two transport regimes can be
distinguished by their shot noise properties.  Our findings for the peak
conductance make it tempting to speculate that the peak conductance will
actually remain at value $e^2/h$ for \emph{any} finite value of the charging
energy which however remains a problem for future studies.

\acknowledgments

We acknowledge fruitful discussions with J.~Stapmanns and financial report via
the Alexander von Humboldt foundation.

\appendix

\section{S-matrix for ideal Majorana bound states}
\label{sec:smatrix_ns}
The scattering matrix for an ideal Majorana bound state at a NS-interface can
be straight-forwardly computed through the Weidenm{\"u}ller formula for the
scattering matrix at energy $\omega$~\cite{nilsson:08}
\begin{align}
s(\omega) = 1 - 2\pi i W^\dagger \frac{1}{\omega/2 + i \pi W W^\dagger} W,
\end{align}
where the vector $W = (w, w^*)$ describes the coupling of the Majorana bound
state to the electron and hole degrees in the lead which is assumed to have
unit density of states.  The scattering matrix can be written as
\begin{align}
s = \begin{pmatrix} s^{ee} & s^{he} \\
	s^{eh} & s^{hh} \end{pmatrix} 
= \begin{pmatrix} 1 + A & A \\
	A & 1+A \end{pmatrix},
\end{align}
with $A = (i\omega/2\Gamma - 1)^{-1}$ where we have introduced the Fermi's
golden rule coupling strength $\Gamma = 2\pi |w|^2$ to the lead.  From this
one obtains the Andreev reflection probability $T^{he} = (1+\omega^2/4
\Gamma^2)^{-1}$.

\end{document}